\documentclass[aps,prb,twocolumn,superscriptaddress,floatfix,longbibliography]{revtex4-2}
\usepackage{graphicx}
\usepackage{dcolumn}
\usepackage{bm}
\usepackage{comment}
\usepackage{amsmath}
\usepackage{xcolor}
\usepackage{float}

\newcommand\norm[1]{\lVert#1\rVert}

\usepackage[colorlinks=true, allcolors=blue]{hyperref}

\begin{document}

\title{Anderson localization: A view from Krylov space}
\author{J. Clayton Peacock}
\email{jcp9552@nyu.edu}
\affiliation{Department of Physics, New York University, New York, NY-10003, USA}
\author{Vadim Oganesyan}
\affiliation{Physics Program and Initiative for the Theoretical Sciences, The Graduate
Center, CUNY, New York, NY 10016, USA}
\affiliation{Department of Physics and Astronomy, College of Staten Island, CUNY,
Staten Island, NY 10314, USA}
\author{Dries Sels}%
\affiliation{Department of Physics, Boston University, Boston, Massachusetts 02215, USA}
\affiliation{Center for Computational Quantum Physics, Flatiron Institute, New York, NY-10010, USA}
\date{\today}

\begin{abstract}
 The Krylov subspace expansion is a workhorse method for sparse numerics that has been increasingly explored as source of physical insight into many-body dynamics in recent years. In this work we revisit the venerable Anderson model of localization in dimensions $d=1, 2, 3, 4$ to construct local integrals of motion (LIOM) in Krylov space. These appear as zero eigenvalue edge states of an effective hopping problem in the Krylov superoperator subspace and can be analytically constructed given the Lanczos coefficients. We exploit this idea, focusing on $d=3$, to study the manifestation of the disorder driven Anderson transition in the anatomy of LIOMs. We find that the increasing complexity of the Krylov operators results in a suppression of the fluctuations of the Lanczos coefficients. As such, one can study the phenomenology of the integrals of motion in the disorder averaged Krylov chain. We find edge states localized on vanishing fraction of Krylov space (of dimension $D_K=V^2$ for cubes of volume $V$), both in localized and extended phases. Importantly, in the localized phase, disorder induces powerlaw decaying dimerization in the (Krylov) hopping problem, producing stretched exponential decay of the LIOMs in Krylov space with a stretching exponent $1/2d$. Metallic LIOMs are completely delocalized albeit across only $\propto \sqrt{D_K}$ states. Critical LIOMs exhibit powerlaw decay with an exponent matching the expected value of $0.29$. 
\end{abstract}

\maketitle

\section{Introduction}
Modeling quenched disorder has been instrumental in understanding the electronic properties of solids. In the case of the non-interacting Anderson model, spatial disorder is known to localize the electronic states if it is larger than a critical disorder strength, whereas below this strength the electron states are extended throughout the system~\cite{Anderson,Mott1968-3d,Kramer1993-localization_review,Evers_2008_AndersonTransitions}. Applying these concepts to interacting systems has been a major endeavor in the past decades~\cite{BASKO20061126,gornyi2005MBL,Oganesyan2007,annurev-conmatphys-031214-014726,Huse2014Phenomenoloy,Imbrie2016,ROS2015420,Serbyn2013,Thomson2018,AbaninReview,annurev-conmatphys-031214-014726,ALET2018498,Schreiber2015,Rubio-Abadal2019,Smith2016,Kohlert2019,Lukin2019,Rispoli2019,Pal2010MBLTransision,Lev2015,Luitz2015,Wahl2017,Mierzejewski2016,Mace2019,Chanda2020,Laflorencie2020,Gray2018,Doggen2018,POLFED2020,Serbyn2015MBLTransition,Sierant2020ThoulessTime,Chanda2020}, but the interacting problem, dubbed many-body localization (MBL), has proved significantly more challenging both conceptually and numerically \cite{suntajs2019quantum,suntajs2020transition,Dynamicalobstruction,sierant2025mbl}. 

We may interpret these localized ergodicity-breaking systems to be effectively integrable, meaning they must possess an extensive set of local integrals of motion (LIOMs).  Their existence may be motivated intuitively, invoking adiabatic evolution of the Hamiltonian from completely decoupled degrees of freedom~\cite{Imbrie2017_LIOMs}, e.g. qubits, in which case LIOMs are Pauli-like operators (i.e. having discrete spectra $\pm 1$), or they can be obtained by simply time-averaging the Hamiltonian evolution of initially localized qubits~\cite{chandran2015constructing,geraedts2017emergent}. Though not uniquely defined, LIOMs should be exponentially localized with a small enough localization length to be stable against so called ``quantum avalanches'' induced by rare Griffith's regions~\cite{Chandran2016_MBLbeyondEigenstates,deRoeck17,Luitz2017_HowaSmallQuantumBathCanThermalize,Ponte2017_HowOneSpin,MorningstarAvalanches,Bath-induceddelocalization,Embedded_Thermal_inclusion,Selsthermalizationofdiluteimpurities,Crowley2020_AvalancheInducedLocalizedandThermalRegions,Potirniche2019_ExplorationofStabilityofMBL,Rubio-Abadal2019,Crowley2022_MeanFieldTheoryofFailedThermalizingAvalanches,Suntajs2022_ErgodicityBreakinginZeroDimensions,Long2023_Prethermal_Phenom,Leonard2023avalanchesexperiment}. Given their non-unique construction and the stringent conditions on their decay, it is important to continue looking for new methods to understand the emergence of local integrals of motion. The Krylov subspace expansion, commonly known as the recursion method \cite{recursionmethod_vis}, is one such method which, although not a new technique, has been increasingly explored as a probe of quantum chaos (for a recent review, see Ref.~\cite{nandy2024quantumdynamicskrylovspace}). The LIOMs we will explore in this work are the same operators as the time-averaged LIOMs of  Ref.~\cite{chandran2015constructing}, but, importantly, expressed in the dynamically generated Krylov basis, which is arguably more natural and demonstrably more compact than any apriori fixed basis.

The edge mode nature of LIOMs studied below has been identified in recent work on nearly integrable spin chains~\cite{Yates2020_LifetimeofAlmostStrong,Yates2021StrongandAlmoststrongModes,Yeh2023SlowlyDecayingZeroMode}.
By way of surveying other recent efforts to exploit Krylov methods, we note there have been many studies of quantum chaos and integrability in Krylov space \cite{Parker2019_universaloperatorgrowth,Barbón2019OnTheEvolutionofOperatorComplexity,Rabinovici2021_OperatorComplexityaJourney,Rabinovici2022Krylovlocalizationandsuppression,Hornedal2022_UltimateSpeedLimitstoGrowthofOpComplexity,Caputa2022_GeometryofKrylovComplexity,Balasubramanian2022_QuantumChaosandtheComplexityofSpreadofStates,Bhattacharjee2022_KComplexity_SaddleDominatedScrambling,Erdmenger2023_UniversalChaoticDynamicsfromKrylov,Rabinovici2022_KrylovComplexityfromIntegrabilitytoChaos,Hashimoto2023_KrylovComplexityandChaos,Suchsland2025_KrylovComplexityandTrotterTransitions,Camargo2024_SpectralandKrylovComplexityBilliards,Cao_2021StatisticalmethodforOG,menzler2024krylovlocalizationprobeergodicity,baggioli2024_krylovcomplexityorderparameter,Yates2020DynamicsofAlmostStrongEdgeModes,Yates2021StrongandAlmoststrongModes,Yeh2023SlowlyDecayingZeroMode,Yeh2024UniversalModelofFloquet,loizeau2025openingkrylovspaceaccess,Trigueros2022_KrylovComplexityofMBL,Balasubramanian2025_ChaosIntegrabilityandLateTimes,Yates2020_LifetimeofAlmostStrong,Scialchi2025_ExploringQuantumErgodicity_KrylovApproach,Uskov2024_QuantumDynamicsinOneandTwoDims,Teretenkov_2025_PseudomodeExpansion,Loizeau_2025_PauliStrings,Yeh2025_MomentMethodandContinuedFractionExpansion,Bartsch2024,Balasubramanian_2025,angelinos2025temperaturedependencekrylovspace} which come with various proposals for detecting ergodicity breaking transitions. For one, chaos has been associated with a maximum growth rate of the Lanczos coefficients \cite{Parker2019_universaloperatorgrowth}, though this is not always the case \cite{Bhattacharjee2022_KComplexity_SaddleDominatedScrambling}. The interplay between integrability, integrability breaking, and dimerization has been pointed out  in Refs. \cite{Rabinovici2022Krylovlocalizationandsuppression,Hashimoto2023_KrylovComplexityandChaos,nandy2024quantumdynamicskrylovspace,menzler2024krylovlocalizationprobeergodicity}. Finally, chaos is associated with delocalization in Krylov space \cite{Dymarsky2020_QuantumChaosAsDelocalization}, which can be quantified by the average position in Krylov space known as the ``Krylov complexity'', the long time average of which is expected to saturate a maximum bound for chaotic systems~\cite{Parker2019_universaloperatorgrowth,Barbón2019OnTheEvolutionofOperatorComplexity,Rabinovici2021_OperatorComplexityaJourney,Rabinovici2022Krylovlocalizationandsuppression,Hornedal2022_UltimateSpeedLimitstoGrowthofOpComplexity,Caputa2022_GeometryofKrylovComplexity,Balasubramanian2022_QuantumChaosandtheComplexityofSpreadofStates,Bhattacharjee2022_KComplexity_SaddleDominatedScrambling,Erdmenger2023_UniversalChaoticDynamicsfromKrylov,Rabinovici2022_KrylovComplexityfromIntegrabilitytoChaos,Hashimoto2023_KrylovComplexityandChaos,Camargo2024_SpectralandKrylovComplexityBilliards}.

In the balance of the paper we introduce the operator Krylov space, generated from a chosen probe operator, and show how the associated Lanczos coefficients can be used to both reproduce that operator's dynamics and construct LIOMs (Section \ref{sec:Krylov}). The Anderson model is introduced in Section \ref{sec:Anderson in Krylov} and we present some summary statistics of its Lanczos coefficients for a particular probe, clearly identifying three different asymptotic decays of zero-modes in the localized phase, critical point, and extended phase. We close (Sec. \ref{sec: conclusions}) with a summary of results and open problems.

\section{Operator Krylov space}\label{sec:Krylov}
We will now briefly describe the Krylov subspace expansion which is explained in greater detail in Refs.~\cite{recursionmethod_vis,nandy2024quantumdynamicskrylovspace}. In the Heisenberg picture operators $O$ evolve in time under the action of a Hamiltonian $H$ as $\frac{d O}{dt} = i[H,O]$, which upon defining the Liouvillian $\mathcal{L}(\cdot) = [H,\cdot]$, allows for the formal solution $O(t)=e^{i\mathcal{L}t}(O)$.  One can do a Taylor expansion of the latter which illuminates the basic idea of the recursion method: to recursively construct a basis of orthogonal operator polynomials. To define these requires defining an inner product on the operator space, and this choice is not unique. Since we're concerned with eigenstate transitions at arbitrary energy, we choose the (normalized) Hilbert-Schmidt norm $||O||^2= D^{-1} {\rm Tr}[O^\dagger O]$, where $D$ is the dimension of the many-body Hilbert space. \par

The recursion method iteratively constructs a Krylov basis through repeated action of the Liouvillian on some initial normalized operator $O_1$ of interest, with associated Lanczos coefficient $b_1=||O_1||=1$. The second element of the basis becomes $O_2=\mathcal{L}(O_1)/\norm{\mathcal{L}(O_1)}=\mathcal{L}(O_1)/b_2$, and the subsequent elements are recursively defined as follows:

\begin{equation}
\begin{split}
    {O'_{n}} & = \mathcal{L}(O_{n-1})-b_{n-1}O_{n-2}\\
    O_{n} & = \frac{O'_{n}}{b_{n}}, \,{\rm with}\, b_n=\norm{O'_{n}}.
    \end{split}
\end{equation}
Written in the Krylov basis $O_n$, the result of the Lanczos algorithm is that $\mathcal{L}$ can now be written in matrix form as:
\begin{equation}\label{eq:L_krylov}
    L_{n,m} = \frac{1}{D} {\rm Tr}[O_n\mathcal{L}(O_m)] = \begin{pmatrix} 
    0 & b_2 & 0  & 0 & \cdots \\
    b_2 & 0 & b_3 & 0 & \cdots \\ 
    0 & b_3 & 0 & b_4 \\
    0 & 0 & b_4 & 0 & \ddots \\
    \vdots & \vdots & & \ddots & \ddots 
    \end{pmatrix}
\end{equation} 
Therefore the Lanczos coefficients $\{b_n\}$, which have units of energy, contain all of the information needed to compute the dynamics of the desired operator $O_1$. The dimension of Krylov space $D_K$ is related to the number of distinct frequencies and thus depends on the model being considered, however it is upper bounded at $D_K \leq D^2 - D + 1$ \cite{Rabinovici2021_OperatorComplexityaJourney}. Systems with level-repulsion are expected to saturate this bound, but some special integrable systems have much smaller $D_K$ \cite{Bhattacharjee2022_KComplexity_SaddleDominatedScrambling} due to their smaller dynamical Lie algebra. In absence of those dynamical symmetries the Krylov space is typically maximal, and as previously noted, the long time limit of the average position in Krylov space, otherwise known as the Krylov complexity, is often used as a measure of integrability~\cite{nandy2024quantumdynamicskrylovspace}).

In addition, as the Lanczos coefficients ${b_n}$ are calculated through repeated action of $n$ powers of $\mathcal{L}$, they are equivalent to knowing the moments of the Liouvillian $\mu_{2n}$ (explained in detail in Ref. \cite{recursionmethod_vis}). The autocorrelation function is also simply defined in this basis as:
\begin{equation}
    C(t) = \frac{1}{D} {\rm Tr}\left[O_1e^{i\mathcal{L}t}(O_1)\right] 
    \label{eq:Correlation_t}
\end{equation}
 Aside from numerical considerations, this construction has conceptual advantage. We can consider $\mathcal{L}$ as a Hamiltonian in its own right, which due to its tri-diagonal nature corresponds to a single particle hopping on a semi-infinite chain, with ${b_n}$ as the hopping amplitudes. In particular, if we express the operator as
\begin{equation}
    O(t) = \sum^{D_{K}}_{n=1}i^{n+1}\phi_n(t)O_n,
\end{equation}
we arrive at the equation of motion 
\begin{equation}
    \frac{d\phi_n}{dt} = -b_{n+1} \phi_{n+1} + b_n \phi_{n-1}.
    \label{eq:phi_t}
\end{equation}

In this work we focus on calculating the integrals of motion, or conserved quantities, in Krylov space. Conserved quantities $Q$ commute with the Hamiltonian, i.e. $\mathcal{L}(Q)=0$, which in Krylov space corresponds to the zero-modes of $L$, which are solutions to expression~\eqref{eq:phi_t} with $\partial_t \phi_n=0$.
These local integrals of motion depend on the initial operator $O_1$. In particular, for a local seed operator $O_1$ and for local Hamiltonians the Krylov expansion is local, i.e. the range of the operator expands by the range of the interactions in the Hamiltonian in each iteration. This offers the interesting possibility to obtain integrals of motion $Q$ that have particularly high overlap with some desired operator $O_1$.

\subsection{Krylov space integrals of motion}\label{sec:Krylov IOMs}

The tri-diagonal structure of $\mathcal{L}$ in the Krylov basis gives a simple solution to the zero-mode as a function of $n$ along the semi-infinite chain in terms of the initial weight on the first site $Q_1$. The support on all the even $n$ vanishes, as they describe currents, while for odd $n$ one arrives at following form (illuminated further in Appendix \ref{App:Zero-modes}):
\begin{equation}
    \begin{aligned}
        Q_{2n+1} =(-1)^{n} \prod_{j=1}^{n
        } \frac{b_{2j}}{b_{2j+1}} Q_1 \,
    \end{aligned}
\end{equation}
or alternatively
\begin{equation}\label{eq:zeromode}
   \frac{|Q_{2n+1}|}{Q_1}= \exp\left(-\sum_{j=1}^n
    \log(b_{2j+1}/b_{2j})\right) \equiv e^{-\kappa(n)}
\end{equation}
Consequently the asymptotic behavior of $\kappa(n)$ controls the properties of the integrals of motion, and it is clear that the dimerization, defined as the ratio of consecutive coefficients, controls their decay. Before moving to the specifics of Anderson localization, let's discuss a few basic constraints and results which will set the stage for what comes. 

In the ideal case, we'd get an exponentially localized zero-mode on the boundary of our Krylov chain, i.e. $\kappa(n) \sim n$. The latter corresponds to a perfectly dimerized chain in which all the odd couplings are greater than the even couplings. This situation essentially describes the Su–Schrieffer–Heeger (SHH) chain \cite{SSH1979} in the topologically non-trivial insulating phase, and our integral of motion corresponds to the well known edge state. While this scenario is realized for the Majorana zero-mode in the transverse field Ising (TFI) chain \cite{Yates2020_LifetimeofAlmostStrong,Yates2021StrongandAlmoststrongModes}, it's extremely fine tuned and unstable to small perturbations \cite{Yeh2023SlowlyDecayingZeroMode}. 

There are some generic properties that highlight the difficulty of finding these types of integrals of motion. First of all, generic seed operators $O_1$ are expected to have matrix elements between all eigenstates of $H$ and we would therefore not expect them to have a spectral gap, e.g. in the eigenstates thermalization regime (ETH \cite{Srednicki1994ETH}) we expect low frequency matrix elements (below the Thouless scale) to behave like random matrix theory. The SSH scenario outlined above is thus highly non-generic as it corresponds to a problem in which our initial operator $O_1$ has a gapped spectrum. Generically, we'd need a scenario where there is localization on the Krylov chain but the spectrum is gapless. Random, or some form of pseudo-random, Lanczos coefficients could give us that~\cite{Fleishman1977_FluctuationsandLocalizationin1D}, but as we'll discuss later this turns out to have other problems. The primary issue is that, while this results in localization, it would be hard to guarantee that the zero-mode is localized close to the boundary. In addition, for local models the Lanczos coefficients are bounded to grow at most linearly $(b_n \sim n)$, with a logarithmic correction in 1D $(b_n \sim n/\log(n))$ \cite{Parker2019_universaloperatorgrowth, Cao_2021StatisticalmethodforOG, nandy2024quantumdynamicskrylovspace}. Typically this growth is saturated and it has the tendency to cause exponentially fast transport on the Krylov chain, further hampering the construction of integrals of motion. 

In order to disentangle the delocalizing effects of many-body operator growth from the localizing effects of disorder, we focus our study on the non-interacting Anderson model. This model has the additional benefit that we can numerically access much larger system sizes. 

\section{Single particle localization in Krylov space} \label{sec:Anderson in Krylov}
We now turn to the Anderson model: a single particle tight-binding model with a disordered potential, given by the following Hamiltonian:
\begin{equation}\label{eq:H_Anderson}
    H = -t\sum_{\langle i,j \rangle}(c^\dagger_ic_j + h.c.) + \frac{W}{2}\sum^{L^d}_{i=1}h_i c^\dagger_ic_i.
\end{equation}
Here $c^\dagger_i,c_j$ are fermionic creation and annihilation operators on sites $i$ and $j$ respectively, the first sum runs over nearest neighbors $\langle i,j \rangle$ with periodic boundary conditions, and the second sum runs over all sites $L^d$ where $d$ is the spatial dimension. The particle has a hopping amplitude $t$, which we set to $t=1$, and random on-site potentials which are constructed from i.i.d. uniformly distributed random numbers $h_i\in[-1,1]$. We refer to $W$ as the disorder strength. In one and two spatial dimensions this model is localized for any $W$, while in three dimensions and higher the model is initially delocalized at low disorder, but undergoes an infinite temperature metal-insulator phase transition at a critical disorder, which is estimated at $W_c \approx 16.5$ in three dimensions \cite{Corrections_to_Scaling_Anderson1999,Statistics_of_Spectra_Anderson1993,SUNTAJS_2021_Spectral_properties_3D} and $W_c \approx 33.2$ in four dimensions \cite{PMarkos_1994_Metal_InsulatorTransition4D}. We will focus on three dimensions in this work, as it is the smallest dimension hosting a transition for which we can explore the extended, critical, and localized regimes with the largest possible linear system size.

\begin{figure}[t]
\includegraphics[width=.45\textwidth]{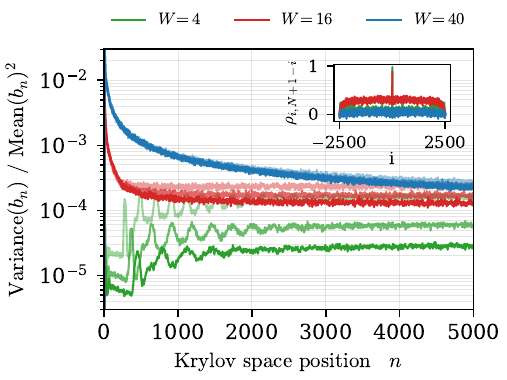}
\caption{Normalized variance of the Lanczos coefficients for the three-dimensional Anderson model at $W=4,16,40$. The inset shows an anti-diagonal cut of the correlation matrix $\rho_{ij}={\rm cov}[b_{i},b_j]/\sigma(b_i)\sigma(b_j)$ which we denote $\rho_{i,N+1-i}$ and which is sharply peaked only at the diagonal index ($i=0$). Different color shades indicate different system sizes: from lightest to darkest are $L=14,18,22$. All results have been obtained from $N_s = 1000$ samples.}
\label{fig:Variances_3d}
\end{figure}

\begin{figure}[t]
\includegraphics[width=.45\textwidth]{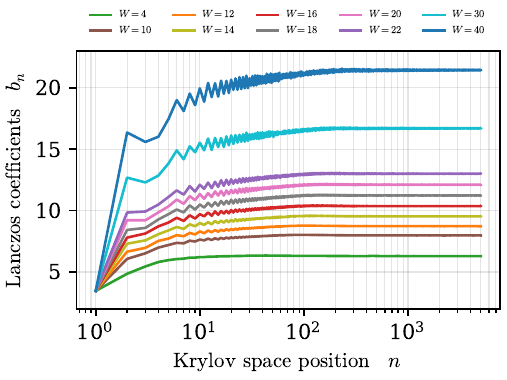}
\caption{Disorder averaged Lanczos coefficients of the three-dimensional Anderson model over a range of different disorders $W$, both above and below the critical disorder (in order from top to bottom are $W=40,30,22,20,18,16,14,12,10,4$). All results are obtained for a system size of $L=22$ with $N_s=1000$ samples.}
\label{fig:LanczosCoefficients_3d}
\end{figure}
Because we are ultimately interested in constructing the conserved quantities, which in the Anderson model are dressed local density operators, we begin our Krylov expansion with a local traceless single site density operator: $O_1 = c^\dagger_xc_x - 1/2$. Here the site $x$ is the site around which we'd like to construct our LIOM, ensuring that the finite amount of Krylov space operators we construct have large initial overlap with that LIOM. We can then write the Krylov expansion of the many-particle problem completely in the single-particle operators, since the entire problem is quadratic in fermionic operators. The single particle Hamiltonian $h$ (an $L^d \times L^d$ matrix) is defined as $H = \sum_{ij}h_{ij}c^\dagger_ic_j$. The single-particle Green function $G$ allows us to express the Heisenberg evolved operators as
    $c_x(t)=\sum_y c_y G(y,t|x,0)=\sum_y c_y \left<y|e^{-iht}|x \right>.$
By Wick's theorem we arrive at following expression for the autocorrelation $C(t)$:
\begin{equation}
    C(t)=\frac{1}{4} |G(x,t|x,0)|^2,
    \label{eq:Coft}
\end{equation}
which is just $1/4$ of the return probability to site $x$. Note that it was important to start with a traceless operator. In conclusion, we can simply perform the Krylov expansion on the single-particle problem by taking a Hilbert-Schmidt inner product on the single particle space and identifying the initial many-body operator $c^\dagger_xc_x-1/2$ with the single particle projector $P_x =1/2 \left|x\right>\left<x\right|$.

We find that in all regimes of the three-dimensional Anderson model the fluctuations are considerably smaller than the means and decrease with $n$ and $L$, as shown in Fig.~\ref{fig:Variances_3d} (the oscillatory behavior for $W=4$ is due to finite size and addressed in Appendix \ref{app:Oscillations}). In addition, the correlation matrix of the $b_n$ is sharply peaked only at the diagonal, so we can conceptually consider the Lanczos coefficients $b_n$ to be become essentially independent random variables $b_n$ with very small fluctuations. Consequently, typical properties of the zero-modes can be understood from the behavior of the disorder averaged Lanczos coefficients. With hindsight this behavior could have been anticipated, since with each increasing order in the Krylov expansion our operator is supported on a larger volume and therefore an increasing amount of random numbers participate in determining the operator structure.

The mean Lanczos coefficients are shown in Fig.~\ref{fig:LanczosCoefficients_3d}, where a clear initial dimerization is observed, which then decreases by some amount. The average $b_n$ across $n$ grows slightly at first but quickly saturates as expected for a free system. In the extended regime (see e.g. $W=10$ in Fig.~\ref{fig:LanczosCoefficients_3d}) this dimerization is very small, but as we will show next, it still affects the form of the zero-mode.

\begin{figure}[t]
\includegraphics[width=.45\textwidth]{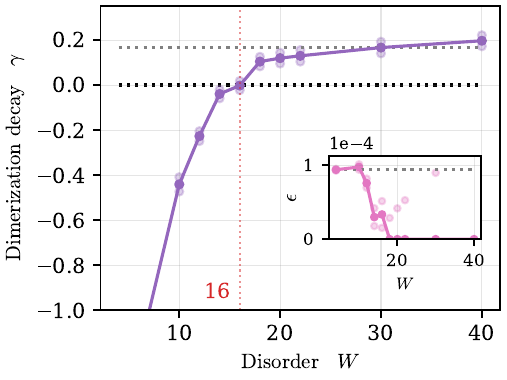}
\caption{Parameters of the dimerization model, expression (\ref{eq:dimerization_model}), fitted by a least squares procedure as a function of disorder (lighter dots show the confidence interval at $10\%$ significance). The critical point of the model, corresponding to $\gamma=0$, occurs at $W_c \approx 16$ consistent with the accepted value $W_c\simeq16.5$. The black dotted line marks the transition point $\gamma=0$ and the grey dotted line marks the localized prediction $\gamma=1/2d$. In the inset, the grey dotted line marks the predicted value in the extended regime: $\epsilon=L^{-d}$}
\label{fig:Dimerization_fit}
\end{figure}

Now that we have established the fluctuations are small, we focus our attention on the disorder averaged coefficients, allowing us to define a simple model for the dimerization of the coefficients:
\begin{equation}\label{eq:dimerization_model}
    \frac{b_{2n+1}}{b_{2n}}\sim 1+\epsilon +\frac{\alpha}{n^{1-\gamma}} 
\end{equation}
The full motivation for the model will become clear throughout the remaining part of the paper. While the model doesn't exactly describe the data, it captures all relevant features. The model is quite accurate as shown in Appendix~\ref{App:dimerization_fitting}. In addition, the parameters are robust as shown by the small error bars in Fig.~\ref{fig:Dimerization_fit}.

\subsection{Weak Disorder} \label{sec: weak disorder}
\begin{figure}[t]
\includegraphics[width=.45\textwidth]{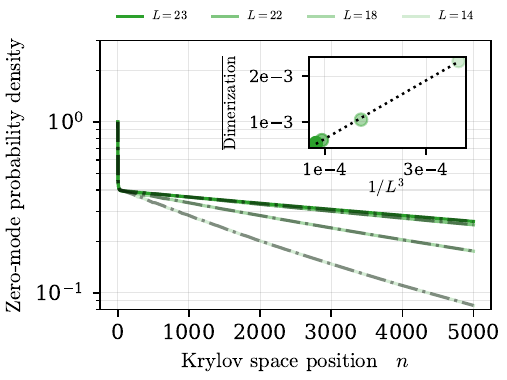}
\caption{Zero-mode probability density in Krylov space for $W=4$ in the extended regime. The dot-dashed line represents the geometric average of the zero-modes constructed in each sample whereas the colored lines are obtained from the sample-averaged $b_n$: the two are in complete agreement. There is an extremely small exponential decay which vanishes as a function of increasing $L$. The inset shows the disorder averaged dimerization of the Lanczos coefficients for different system sizes $L=14,18,22,23$, revealing its inverse scaling with the volume roughly as $\simeq 6.3L^{-3}$. There are $N_s=1000$ samples for all sizes except $L=23$ for which there are $N_s=500$.}
\label{fig:Extended_regime}
\end{figure}

In the extended regime we might expect the zero-mode to be delocalized in Krylov space. However, careful analysis of this zero-mode, shown in Fig.~\ref{fig:Extended_regime}, reveals that the zero-mode is actually exponentially localized albeit extremely weakly. While this might at first sight seem to contradict our earlier statements, it's perfectly consistent. Recall that exponential localization is associated with perfect dimerization of the Lanczos coefficients and hence with a spectral gap. Since the numerics are performed on a finite size system $L$ and we reach Krylov space indices which far exceed this linear system size, we are actually picking up on the finite size gap. On the delocalized side, the system is chaotic and thus exhibits level repulsion. As such, the typical smallest level spacing is of $O(1/V)$. As shown in Fig.~\ref{fig:Extended_regime} the average dimerization, and thus the associated Krylov localization length, goes to zero in the thermodynamic limit as $1/L^3$.  This is confirmed with our fit of expression \ref{eq:dimerization_model} in the inset of Fig. \ref{fig:Dimerization_fit} where $\epsilon \approx 1/L^3 \approx 9\times10^{-5}$.

While this makes sense from the perspective of the spectral finite-size effects, note that the total size of Krylov space is quadratically larger $D_K=O(V^2)$. Over the entire range from $O(V)$ to $O(V^2)$ the zero-mode is thus exponentially small. That this is inevitable also follows from a simple mathematical constraint. Consider again the two-time correlation function given by expression~\eqref{eq:Correlation_t}. In Krylov space, the long time limit of the correlation function in expression~\eqref{eq:Coft} is simply given by the zero-mode probability on the first site: 
\begin{equation}
    C_\infty=\lim_{t \rightarrow\infty} C(t)=|Q_1|^2.
\end{equation}
The latter simply follows from inserting the eigenbasis of $L_{n,m}$ and assuming that there is only a single zero eigenvalue state. The same quantity can of course be calculated in the underlying Hamiltonian problem by taking the infinite time-average of $C(t)$, which results in
\begin{equation}
    C_\infty=\frac{1}{D}\sum_{n=1}^D |\langle n|O_1|n \rangle|^2,
    \label{eq:CinftyH}
\end{equation}
where $ \left| n \right>$ denote the eigenstates of the many-body Hamiltonian, i.e. $H  \left| n \right>=E_n  \left| n \right>$. Both expressions should result in the same answer, and consequently expression~\eqref{eq:CinftyH} constrains $|Q_1|^2$. For a system with conserved particle number that is otherwise ergodic, one would expect~\eqref{eq:CinftyH} to be of $O(1/V)$. The latter explains why the zero-mode needs to be localized in Krylov space on the first $O(V)$ sites. Now we can be slightly more specific. Since our theory is non-interacting, we can recast~\eqref{eq:CinftyH} in terms of single particle eigenstates $\psi_n(x)$. Recall our initial operator $O_1=c^\dagger_xc_x-1/2$, such that one arrives at: 
\begin{equation}
    C_\infty=\sum_{n=1}^V |\psi_n(x)|^4 \geq \frac{1}{V}.
\end{equation}
In the delocalized phase we thus have $C_\infty=O(1/V)$ by definition. In addition, the inequality shows that it's simply not possible for the Krylov zero-mode to be delocalized over the full Krylov space. 

\subsection{Critical Disorder} \label{sec: crit disorder}

\begin{figure}[t]
\includegraphics[width=.45\textwidth]{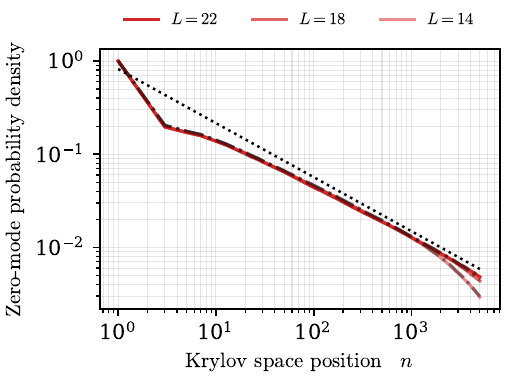}
\caption{The zero-mode probability density for the three-dimensional Anderson model in Krylov space close to the critical point ($W=16$) is shown to decay as a powerlaw $n^{-2\alpha}$ with an exponent $\alpha$ approaching $\alpha\approx0.29$ (shown by the dotted line). The dot-dashed line represents the geometric average of the zero-modes constructed in each sample whereas the colored lines are obtained from the sample-averaged $b_n$: the two are in complete agreement.}
\label{fig:Critical_LIOM}
\end{figure}

The Krylov space properties of the critical point can be constrained by the same analysis we've conducted on the delocalized side. The Anderson critical point is known to exhibit multi-fractality \cite{Janssen1994_Multifractal_localization,Evers_2008_AndersonTransitions} and was recently studied in detail in Ref.~\cite{Hopjan2023_Scale_InvariantSurvivalProb}. This multi-fractality manifests itself as powerlaw scaling of the survival probability $C(t)\sim t^{-\beta}$, with $\beta \approx 0.42$. At the critical point, the Thouless time approaches the Heisenberg time (and is $O(V)$) implying  
\begin{equation}
    C_\infty \sim V^{-\beta}.
\end{equation}
Consequently, the zero-mode must itself decay like a powerlaw:
\begin{equation}
    |Q_n| \sim \frac{1}{n^\alpha}.
\end{equation}
Since the finite size gap, i.e. the inverse Heisenberg time, is still $O(1/V)$, one immediately finds that $\beta=1-2 \alpha$ and $\alpha \approx 0.29$. This scaling is indeed numerically observed, as shown in Fig. \ref{fig:Critical_LIOM}, where for increasing $L$ the zero-mode is approaching the expected power-law.

These conclusions follow from rather general physical constraints. At this point, it's interesting to tie this back to the behavior of the Lanczos coefficients. A powerlaw decaying zero-mode implies (c.f. expression~\eqref{eq:zeromode}) that $\kappa(n) \sim \alpha \log(n)$. Consequently, the Lanczos coefficients asymptotically behave like:
\begin{equation}
    \frac{b_{2n+1}}{b_{2n}}\sim 1+\frac{\alpha}{n}
\end{equation}
This is consistent with the transition point of our model: $\gamma=0$. If the dimerization would decay any faster than $1/n$, it would result in delocalization, whereas slower decay would result in a localized zero-mode.

\subsection{Strong Disorder} \label{sec: strong disorder}

Since eigenstates are localized at strong disorder, $C_\infty$ needs to be finite in the thermodynamic limit. As such, the zero-mode needs to be a square normalizable wavefunction on the half-chain. The only natural way to continuously connect such scaling across the phase transition is to assume that the dimerization behaves like a powerlaw, and therefore we arrive again at our dimerization model for the coefficients, with $\epsilon=0$ since the localized regime is not sensitive to the finite size.
\begin{equation}
    \frac{b_{2n+1}}{b_{2n}}\sim 1+\frac{\alpha}{n^{1-\gamma}} \nonumber
\end{equation}
As discussed, $\gamma=0$ at the critical point and $\gamma<0$ on the delocalized side. Such scaling would result in stretched exponential zero-modes with 
\begin{equation}
    \kappa(n) \sim n^\gamma \nonumber
\end{equation}
The only remaining question is how the stretching exponent $\gamma$ behaves on the localized side. We already know that $\gamma=0$ at the critical point and $\gamma<1$ in order for the spectrum to remain gapless. Given that the localization length fully captures the behavior of the quantum states in the localized phase, it's not obvious by what mechanism $\gamma$ would depend on the disorder strength $W$. 

As such, we suggest that $\gamma$ is fixed to some value $0<\gamma <1$ deep in the localized phase. Recall that, both at the critical point and in the delocalized phase, we mapped real space properties to Krylov space by some rather generic arguments. Proceeding in the same vein, the absence of level repulsion in the delocalized phase suggests that the zero-mode can now span the entire Krylov space of size $O(V^2)$. At the same time, the only relevant length scale is the localization length $\xi$, implying the only dimensionless parameter is $L/\xi$. As such, one simply expects Krylov space to be some rescaled version of real space, i.e. $[n]^{1/2d}=L$ where $d$ is the number of spatial dimensions of the model. 

\begin{figure}[t]
\includegraphics[width=.45\textwidth]{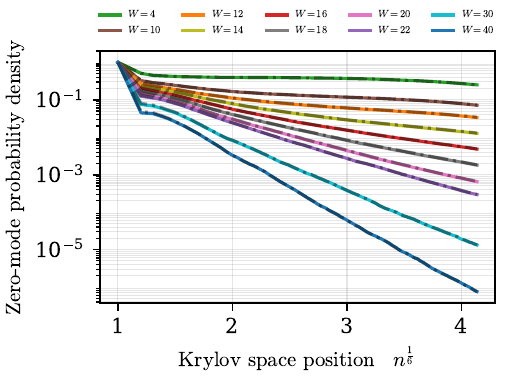}
\caption{Zero-mode probability density for the three-dimensional Anderson model in Krylov space approaches a stretched exponential decay as $W$ is increased into the localized regime ($W>16$). The results are shown as a function of $n^{1/2d}$ (with $d=3$), the predicted stretching coordinate deep in the localized phase. The dot-dashed line represents the geometric average of the zero-modes constructed in each sample whereas the colored lines are obtained from the sample-averaged $b_n$: the two are in complete agreement.}
\label{fig:LIOMs-all-3d}
\end{figure}

\begin{figure}[t]
\includegraphics[width=.45\textwidth]{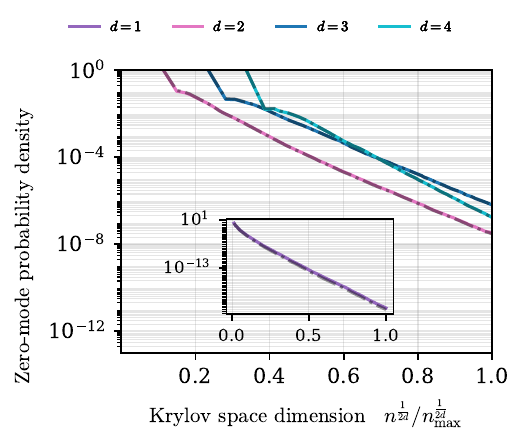}
\caption{The zero-mode probability density in Krylov space for Anderson models in different spatial dimensions $d$. In all cases the disorder is chosen to be much greater than the critical disorder such that all states are well localized. The results are shown as a function of the anticipated stretched Krylov coordinate $n^{1/2d}$, rescaled by the maximum value $n_{\text{max}}^{\frac{1}{2d}}$ so that all curves appear in the same scale. $d=1$ appears in the inset which has the same axes but an extended vertical range. The dot-dashed line represents the geometric average of the zero-modes constructed in each sample whereas the colored lines are obtained from the sample-averaged $b_n$: the two are in complete agreement. Results for spatial dimensions $d=1,2,3,4$ have disorders $W=10,20,40,80$, lattice lengths $L=12000,110,23,10$, and number of samples $N_s = 500,1000,1000,1000$ respectively.}
\label{fig:LIOMs-all-d}
\end{figure}

We thus propose that deep inside the localized phase the dimerization decays with
\begin{equation}
    \gamma=\frac{1}{2d}.
\end{equation}
This results in a stretched exponential zero-mode with a probability density given by $|Q_n|^2 \sim \exp{(-cn^{1/2d})}$. This behavior is consistent with our numerical data for $W>16$, as shown in Fig.~\ref{fig:LIOMs-all-3d}, and is additionally confirmed by our fit of the dimerization model in Fig.~\ref{fig:Dimerization_fit} where $\gamma= 1/2d \approx 0.167$ is found consistently in the localized regime. To further substantiate our proposed scaling, we repeat the same analysis for Anderson models ranging from one to four spatial dimensions. The results shown in Fig.~\ref{fig:LIOMs-all-d} all indicate perfect stretched exponential decay with the predicted exponent $\gamma$ (Lanczos coefficients for these results are presented in Fig. \ref{Fig:bn_alld} in Appendix \ref{App:coefficients_alld}.

\section{Conclusions and Outlook}\label{sec: conclusions}
In summary, we have described the form of LIOMs in the Anderson model in Krylov space as generated by projectors on single sites. There are three types of behaviors -- stretched exponential decay inside the localized phase (with geometric exponent $\gamma=1/2d$ set by real space dimension $d$), delocalization on real-space volume $V=\sqrt{D_K}$ states in the metallic/extended phase, and powerlaw decay at the critical point. Importantly, these results may be interpreted using effective disorder-free chains obtained from disorder averaged Krylov coefficients, which allows considerable further insight into the Anderson transition: e.g. the significance of Krylov dimerization as a mechanism of localizing operators despite having a gapless Liouvillian. The critical point occurs when the average dimerization decreases faster than a powerlaw with power 1, which is the point at which the zero-modes cease to be normalizable in Krylov space ($\gamma=0$ in expression \eqref{eq:dimerization_model}). The dimerization is thus a useful probe to discriminate between ergodic and non-ergodic phases when the average operator growth is otherwise the same.

Our work may be extended to other aspects of the Anderson transition: for example, to perhaps enable the tracking of the mobility edge as a function of typical energy. It also sets up a framework for extending these results to interacting models where the coefficients grow with some power $b_n\sim n^\delta$ (with $\delta \leq1$). If the dimerization decays like a powerlaw -- since only the ratio of Lanczos coefficients matters for the zero-mode -- the growth simply re-normalizes that power. Consequently, the faster the coefficients grow, i.e. the faster the operator growth, the slower the dimerization must decay in order for conserved quantities to exist. For spin systems debated to show MBL, such as the disordered Heisenberg model, the coefficients have been found to grow as $\sim n/\log{n}$ on average, with some dimerization \cite{Trigueros2022_KrylovComplexityofMBL,Weisse2025_OperatorGrowthMBL}. The $\log$ correction would only be relevant at the critical power implying that the dimerization has to be constant for conserved quantities to exist. Interestingly, this still leaves the possibility for the dimerization to decay in interacting integrable models such as the Heisenberg model, where there is typically square root growth \cite{Parker2019_universaloperatorgrowth}.

\emph{Data Availability} -- Basic Julia code for generating the Lanczos coefficients and constructing the local integrals of motion in the Anderson model can be found at \url{https://github.com/cpeacockc/Anderson-in-Krylov-space}

\emph{Acknowledgements} -- We acknowledge useful discussions with Hsiu-Chung Yeh, Aditi Mitra, Anatoli Polkonikov, and David Long. 
The Flatiron Institute is a division of the Simons Foundation. D.S. and J.C.P. thank AFOSR for support through Award no. FA9550-25-1-0067.

\bibliography{bib}

\appendix
\section{Solving for the zero-modes in Krylov space}\label{App:Zero-modes}
We will briefly illuminate the derivation of expression \eqref{eq:zeromode}, which is found by solving $L_{n,m}Q_n =0$ for $Q_n$ in Krylov space:
\begin{equation}
    L_{n,m} Q_n = \begin{pmatrix} 
    0 & b_2 & 0  & 0 & \cdots \\
    b_2 & 0 & b_3 & 0 & \cdots \\ 
    0 & b_3 & 0 & b_4 \\
    0 & 0 & b_4 & 0 & \ddots \\
    \vdots & \vdots & & \ddots & \ddots 
    \end{pmatrix} \begin{pmatrix}
        Q_1 \\ Q_2 \\ Q_3 \\ Q_4 \\ \vdots
    \end{pmatrix} = \begin{pmatrix}
        0 \\ 0 \\ 0 \\ 0 \\ \vdots
    \end{pmatrix}
\end{equation} 
Then, one can directly compute:
\begin{align*}
    Q_2b_2 &= 0 \quad \& \quad Q_3 = -\frac{b_2}{b_3} Q_1 \\
    Q_4b_4 &= 0 \quad \& \quad Q_5 = -\frac{b_4}{b_5} Q_3\\
    &\vdots \quad \quad \quad \quad \quad \quad  \vdots \\
    Q_{2n} &=0 \quad \& \quad Q_{2n+1} =(-1)^{n} \prod_{j=1}^{n
        } \frac{b_{2j}}{b_{2j+1}} Q_1,
\end{align*}
arriving at expression \eqref{eq:zeromode} for the Krylov space zero-mode.

\section{Variance oscillations in the extended regime variances}\label{app:Oscillations}

\begin{figure}[h]
\includegraphics[width=.45\textwidth]{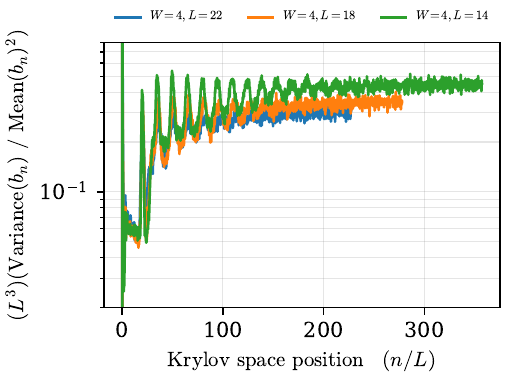}
\caption{Normalized and rescaled variance of the Lanczos coefficients in the extended regime ($W=4$) for system sizes $L=14,18,22$ (top to bottom). The oscillations are due to finite size recurrences, which is shown by their collapse when rescaling the Krylov space by the linear system size $(n \rightarrow n/L)$. The vertical axis is rescaled by the volume $(\times L^3)$ to better show the collapse.}
\label{Fig:Variances_W4}
\end{figure}

In Fig. \ref{fig:Variances_3d} there are oscillations in the extended regime due to finite size recurrences due to the small linear system sizes. This is clearly seen in Fig. \ref{Fig:Variances_W4} where the Krylov space has been rescaled by $n \rightarrow n/L$, showing a collapse of the oscillatory frequencies.

\section{A closer look at the dimerization}
\begin{figure}[H]
\includegraphics[width=.45\textwidth]{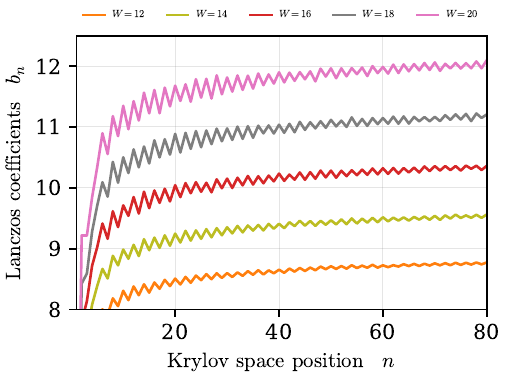}
\caption{Lanczos coefficients for $L=22$ in three dimensions and disorders (from top to bottom) $W=20,18,16,14,12$}
\label{Fig:Dimerization}
\end{figure}

In Fig. \ref{Fig:Dimerization} we show the Lanczos coefficients for a few disorders, zoomed in so the dimerized pattern is more clear. This pattern is very stable (i.e. even coefficients are larger than surrounding odd ones), and one can see by eye a reduction in the dimerization across the critical point.

\section{The dimerization fitting procedure}\label{App:dimerization_fitting}
\begin{figure}[ht]
\includegraphics[width=.45\textwidth]{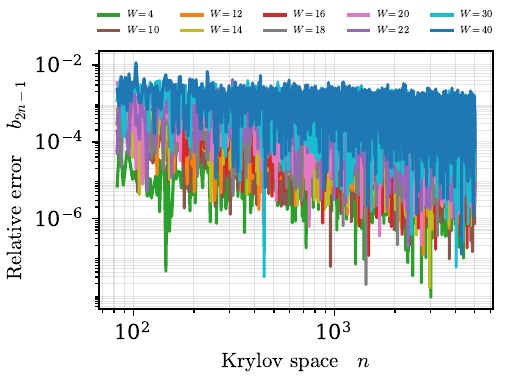}
\caption{Relative errors for the odd sample-averaged Lanczos coefficients shown in Fig. \ref{fig:LanczosCoefficients_3d} using the fitted parameters shown in Fig. \ref{fig:Dimerization_fit}.}
\label{Fig:fit_errors}
\end{figure}

The parameters in Fig. \ref{fig:Dimerization_fit} were found by fitting the sample-log-averaged dimerization $(\frac{b_{2n+1}}{b_{2n}}-1)$  for $n>40$ with a least squares routine following the model defined in expression $\eqref{eq:dimerization_model}$. The dimerization was constructed in each sample first before taking the average. Within the fit, $\epsilon$ was set to be strictly non-negative, while $\gamma$ and $\alpha$ were unrestricted. As a proof of concept of this fit, we used the fitted parameters to predict the odd sample-averaged Lanczos coefficients $b_{2n+1}$ from the even ones $ b_{2n}$, and plot the subsequent relative error in Fig. \ref{Fig:fit_errors}, which is shown to be both less than $10^{-2}$ across Krylov space for all disorders. 

\section{Lanczos coefficients in all spatial dimensions}\label{App:coefficients_alld}
\begin{figure}[ht]
\includegraphics[width=.45\textwidth]{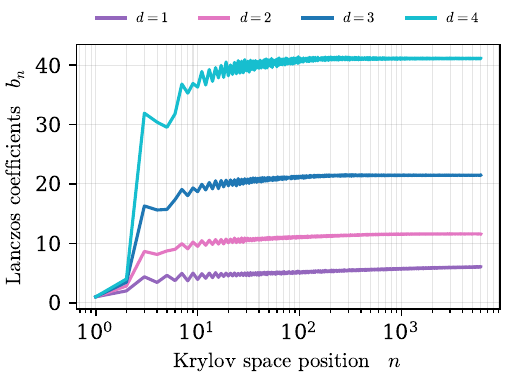}
\caption{Lanczos coefficients for spatial dimensions $d=1,2,3,4$ (top to bottom) with respective disorders $W=10,20,40,80$, lattice lengths $L=12000,110,23,10$, and number of samples $N_s = 500,1000,1000,1000$ respectively.}
\label{Fig:bn_alld}
\end{figure}

Here in Fig. \ref{Fig:bn_alld} we show the Lanczos coefficients for spatial dimensions $d=1,2,3,4$ as used to construct the zero-modes in the respective localized phases in Fig. \ref{fig:LIOMs-all-d}.

\end{document}